\documentclass[11pt]{article}
\usepackage{euscript}
\usepackage{amssymb}
\usepackage{amsfonts}
\usepackage{amsbsy}
\usepackage{amsmath}
\usepackage{epsfig}
\usepackage{amsthm}
\usepackage{amscd}
\usepackage{amstext}
\usepackage{verbatim}

\usepackage[pdftex]{hyperref}

\def\ben{\begin{equation}}
\def\een{\end{equation}}
\def\half{{\textstyle{1\over2}}}

    \let\p=\phi

\let\w=\omega

\def\be{\begin{equation}}
\def\ee{\end{equation}}
\def\beq{\begin{equation}}
\def\eeq{\end{equation}}
\def\ba{\begin{array}}
\def\ea{\end{array}}

\def\dalemb#1#2{{\vbox{\hrule height .#2pt
       \hbox{\vrule width.#2pt height#1pt \kern#1pt
               \vrule width.#2pt}
       \hrule height.#2pt}}}

\newcommand{\bea}{\begin{eqnarray}}
\newcommand{\eea}{\end{eqnarray}}
\newcommand{\tr}{{\rm tr} }
\newcommand{\Tr}{{\rm Tr} }

\renewcommand{\p}{\partial}

\newcommand{\tP}{\tilde{P}}
\newcommand{\tL}{\tilde{L}}
\newcommand{\tQ}{\tilde{Q}}

\begin{document}

\begin{center}

{ \LARGE {\bf Warped Conformal Field Theory}}

\vspace{1.2cm}

St\'{e}phane Detournay$^\sharp$, Thomas Hartman$^\flat$ and Diego M. Hofman$^\sharp$$^*$ 

\vspace{0.9cm}

{\it $^\sharp$ Center for the Fundamental Laws of Nature, Harvard University,\\
Cambridge, MA 02138, USA \\}

\vspace{0.5cm}

{\it $^\flat$ Institute for Advanced Study, School of Natural Sciences, \\
Princeton, NJ 08540, USA \\}

\vspace{0.5cm}

{\it $^*$ Theory Group, SLAC National Accelerator Laboratory \\
Menlo Park, CA 94025, USA \\}

\vspace{1.6cm}

{\tt sdetourn@physics.harvard.edu, hartman@ias.edu, dhofman@slac.stanford.edu} \\

\vspace{1.6cm}

\end{center}

\begin{abstract}
We study field theories in two spacetime dimensions invariant under a chiral scaling symmetry that acts only on right-movers.  The local symmetries include one copy of the Virasoro algebra and a U(1) current algebra.  This differs from the 2d conformal group, but in some respects is equally powerful in constraining the theory.  In particular, the symmetries on a torus lead to modular covariance of the partition function, which is used to derive a universal formula for the asymptotic density of states.  For an application we turn to the holographic description of black holes in quantum gravity, motivated by the fact that the symmetries in the near horizon geometry of any extremal black hole are identical to those of a 2d field theory with chiral scaling.  We consider two examples: black holes in warped AdS$_3$ in topologically massive gravity, and in string theory.  In both cases, the density of states in the 2d field theory reproduces the Bekenstein-Hawking entropy of black holes in the gravity theory.

\end{abstract}

\pagebreak
\setcounter{page}{1}

\tableofcontents

\pagebreak

\section{Introduction}

The structure of conformal field theories (CFTs) in two spacetime dimensions is rich enough to determine many properties of the underlying theories. An important example is the number of states at high energy, which is fixed by conformal symmetry and depends only on the central charges in a unitary theory. This result is especially powerful given that in two dimensions, scale invariance implies conformal invariance:  any unitary theory with a discrete spectrum and invariant under two dimensional translations, Lorentz transformations and scalings has an enlarged global symmetry group given by $SL(2, R) \times SL(2,R)$ and local symmetries given by two copies of the Virasoro algebra \cite{Polchinski:1987dy}. If we add the requirement of modular invariance for consistency on a torus, we obtain the famous Cardy result for the entropy of a CFT \cite{Cardy:1986ie}:
\be\label{cardy0}
S_{\rm CFT} = 2\pi\sqrt{\frac{c_R}{6}L_0} + 2\pi\sqrt{\frac{c_L}{6}\bar{L}_0} \ .
\ee

This universal result helped uncover the deep connection of CFTs to
black hole microphysics, a manifestation of gauge/gravity duality, or more specifically the AdS/CFT 
correspondence \cite{Maldacena:1997re, Witten:1998qj, Gubser:1998bc}. The entropy of asymptotically AdS$_3$ black holes, as well as that of higher-dimensional black holes with an AdS$_3$ near-horizon geometry, can be calculated by identifying two copies of the centrally-extended Virasoro algebra in the asymptotic symmetries \cite{Brown:1986nw}. The states of the corresponding quantum theory, if it exists, must form 
representations of that algebra, hence the theory is a 2d CFT. Using the Cardy formula (\ref{cardy0}) to count the degeneracy of states at high energy reproduces the Bekenstein-Hawking area law for the black hole entropy \cite{Strominger:1997eq}. Though striking, this derivation does not allow for a precise identification of the corresponding microscopic degrees of freedom, although this can be achieved by a detailed string theory treatment in special circumstances \cite{Strominger:1996sh, Maldacena:1997de}. 

It is, of course, of interest to extend holography to other spacetimes that are not asymptotically AdS. Much effort has been dedicated to the study of flat (see \cite{Banks:1996vh,Polchinski:1999ry, Barnich:2006av, Bagchi:2010eg, Barnich:2010eb, Bagchi:2012cy, Bagchi:2012yk,Bagchi:2012xr, Barnich:2012xq}  for example) and de Sitter (see e.g. \cite{Hawking:2000da,Strominger:2001pn,Alishahiha:2004md,Dong:2010pm,Maldacena:2011mk,Anninos:2011ui,Anninos:2011af}) spacetimes, of clear importance for physical applications.  Interesting cases studied recently in connection with condensed matter applications are Lifshitz geometries \cite{Kachru:2008yh,Gonzalez:2011nz} and hyperscaling violating geometries \cite{Huijse:2011ef}. Unfortunately, in most of these cases, little is known about the putative dual field theory, so it is of interest to find non-AdS examples where we have more information about the structure of the dual.

In this paper we consider 2d QFTs with a chiral scaling symmetry that acts only on right-movers, $x^- \rightarrow \lambda x^-$. In contrast to CFTs, which have a second, independent scaling symmetry $x^+ \rightarrow \bar{\lambda}x^+$, we require only translation invariance on the left. A field theoretic study of these symmetries was performed in \cite{Hofman:2011zj}, generalizing the results of \cite{Polchinski:1987dy} and leading to the following conclusion: a 2d translation-invariant theory with a chiral scaling symmetry must have an extended local algebra.  There are two minimal options for this algebra.  One is the usual CFT case with two copies of the Virasoro algebra.  The other possibility is one Virasoro algebra plus a $U(1)$ Kac-Moody algebra.\footnote{While holomorphic (ie, chiral) CFTs containing a current exhibit this algebra, the $U(1)$ does not correspond to a spacetime translation. We'll have more comments to make about the connection between these theories below, but they are inequivalent, and the theories considered here do not need to be holomorphic.} We call this case a Warped Conformal Field Theory (WCFT). 

In the first part of this paper, we study these theories from a purely field theoretic viewpoint.  The symmetries impose powerful constraints on the theory, much like in CFT.  This may seem surprising, because the global symmetries of a WCFT are $SL(2,R)\times U(1)$, a subset of those in CFT.  However, the local symmetries include two arbitrary functions worth of freedom in coordinate transformations:
\be
x^- \rightarrow f(x^-) \ , \quad\quad x^+ \rightarrow x^+ - g(x^-) \ .
\ee
These symmetries are used to derive a new type of modular transformation on the torus.  Applied to the finite-temperature partition function, the modular transformation relates thermodynamic quantities at slow rotation to those at high rotation.  This leads to constraints on the spectrum of a WCFT, and a universal result for the asymptotic entropy: 
\be\label{wcardy0}
S_{\rm WCFT} = -\frac{4\pi i P_0 P_0^{vac}}{k} + 4\pi\sqrt{-\left(L_0^{vac} - \frac{(P_0^{vac})^2}{k}\right)\left(L_0 - \frac{P_0^2}{k}\right)} \ .
\ee
Here $L_0$ is the charge associated to the $ SL(2,R)$ zero mode, $P_0$ is the $U(1)$ charge, $c$ and $k$ are the central extensions of the Virasoro + Kac-Moody algebra, and `vac' labels the vacuum state. This is analogous to the Cardy formula (\ref{cardy0}) in an ordinary CFT, but is not the same and does not rely on conformal invariance. While the $i$ in the formula above might seem puzzling, we will see that in the examples we will consider $S$ is manifestly real.

Despite this universal result, little is known about these theories in detail, and no nontrivial field theory example is known to have a chiral scaling symmetry.\footnote{A trivial example might be constructed by the relevant deformation of a CFT by a current operator. Notice, however, that this deformation is topological and, therefore, does not change the local physics.} For this reason, in the second part of the paper when we consider examples we will focus on holographic constructions of WCFTs, which do exist. The global symmetries of a WCFT, $SL(2,R) \times U(1)$, are precisely the isometries that appear in the near horizon geometry of every extremal black hole, in any number of dimensions.  This hints at a role for WCFTs in the holographic description of black holes beyond the realm of AdS.

We will restrict ourselves to the specific example of black holes in warped AdS$_3$ (WAdS$_3$). This spacetime is a deformation of AdS$_3$ that changes the asymptotics but preserves $SL(2,R)$ $\times U(1)$ isometries, so it is a simple testing ground for the holographic construction of WCFTs.  Warped AdS is also ubiquitous in extremal black hole geometries; for example, the near horizon geometry of the extremal 4d Kerr black hole at fixed polar angle is WAdS$_3$.  This spacetime is non-Einstein, so one of the simplest theories in which WAdS$_3$ appears is topologically massive gravity (TMG), 3d gravity with a gravitational Chern-Simons term.  Black holes in this theory have been constructed, and their entropy calculated \cite{Nutku:1993eb,GursesBH,Bouchareb:2007yx,Moussa:2003fc}. The result is surprising, as it matches the Cardy formula, even though the full conformal symmetry is not apparent \cite{Anninos:2008fx}. 

Further investigation into this matter led to the calculation of the asymptotic symmetry group (ASG) of this spacetime \cite{Compere:2008cv,Blagojevic:2008bn,Compere:2009zj,Blagojevic:2009ek,Henneaux:2011hv}. The result in general may depend on the choice of boundary conditions, but for the only choice that is known to be consistent, it was shown that the symmetries consist not of two copies of the Virasoro algebra, but of one Virasoro algebra and a 
$U(1)$ current algebra extending the exact isometries. This suggests that the dual theory to WAdS$_3$, if it exists, should exhibit these symmetries, seemingly at odds with the proposal in \cite{Anninos:2008fx} that the dual is a CFT. In stringy realizations of WAdS$_3$, worldsheet results corroborate the ASG analysis \cite{Detournay:2010rh,Azeyanagi:2012zd}. Apparently, if a second Virasoro algebra exists, then it must be hidden in the standard representation of the bulk fields \cite{Guica:2011ia}.

On the field theory side, it was argued that the holographic duals to these theories can be constructed by flowing to the IR of a dipole-deformed field theory \cite{ElShowk:2011cm,Song:2011sr}, introducing a degree of non-locality in the picture.  However, it is unknown how to characterize the theory in the IR.

We will take a complementary approach, exploring the conjecture that the dual field theory is a WCFT.  This is not a microscopic definition of the field theory, but because of the powerful constraints imposed by WCFT symmetries, it does allow nontrivial checks.  In particular, we show that (\ref{wcardy0}), the universal entropy formula of WCFT, equals the Bekenstein-Hawking entropy of warped black holes in TMG and in a stringy realization of WAdS$_3$.

Perhaps this can be used to shed light on the proposed Kerr/CFT correspondence, in which extremal black holes are related to a 2d CFT \cite{Guica:2008mu} (see \cite{Compere:2012jk} for a recent review). In that case the Cardy formula (\ref{cardy0}) reproduces the black hole entropy, despite the fact that the $SL(2,R) \times SL(2,R)$ global symmetry is absent, and modular invariance has no obvious bulk analogue.  Although we will not address Kerr directly, we compare the WCFT entropy formula to the Cardy formula and show how they are naturally related by a non-local reparameterization of the WCFT algebra. This partially resolves the puzzle mentioned above, that in warped AdS, TMG behaves like a CFT \cite{Anninos:2008fx}, while the readily available symmetries are those of a WCFT. It also resolves a second puzzle in TMG, that appears for Kerr as well:  matching to the Cardy formula in the microcanonical ensemble requires seemingly arbitrary shifts in the zero mode charges. In the WCFT picture, these shifts are fixed unambiguously.

It is worth mentioning, in passing, that this type of approach might prove useful in the analysis of higher spin theories. Here the presence of more general conserved currents forces us to study the problem in a way in which they are all considered equally. In fact, a WCFT can be viewed as a sort of geometrization of $U(1)$ current algebra; in higher spin theory, gauge currents and geometry are also mixed by gauge transformations and should be treated on an equal footing. This type of approach may be a useful way to understand the modular properties of partition functions in these theories, as studied in \cite{Kraus:2011ds, Gaberdiel:2012yb}.

The layout of this paper is the following. In section 2 we give a precise definition of what we mean by a Warped Conformal Field Theory. We discuss its algebra and unitary representations. Furthermore we show how the currents transform under finite changes of coordinates. In section 3 we discuss the partition functions of these theories and show they transform nicely under modular transformations. We use this result to obtain an expression for the entropy at large values of the charges (i.e. the asymptotic density of states). In section 4 we discuss a slight generalization of this result to other ensembles and relate this to a nonlocal modification of the WCFT symmetry algebra. It turns out this is the relevant framework to understand certain examples of WCFTs that appear as holographic duals of three dimensional gravitational models and string theory. In section 5 we study the specific case of Topologically Massive Gravity. We review the relevant results from the literature and show that the Bekenstein-Hawking entropy of these theories is reproduced by our general expression. Furthermore, we argue these theories cannot be unitary (in the range of couplings we consider and with the standard boundary conditions). In section 6 we discuss a better behaved example coming from string theory. In particular we discuss the example recently discussed in \cite{Azeyanagi:2012zd} obtained by TsT transformations of AdS$_3 \times S^3$ NS-NS backgrounds \cite{Song:2011sr}. Section 7 provides a discussion of our results. Finally, appendix A outlines the derivation of the Cardy formula in ordinary CFT while appendix B sets conventions by defining charges in TMG.

\section{Symmetries}\label{s:primer}

We take as a definition of a Warped Conformal Field Theory (WCFT) the nontrivial minimal case corresponding to the symmetry structure present in a 2d Lorentzian theory with $SL(2,R)_R \times U(1)_L$ global invariance. The results obtained in \cite{Hofman:2011zj} imply the existence of both a right moving energy momentum tensor and a right moving $U(1)$ Kac Moody current. The zero modes generate the global $SL(2,R)_R$ and $U(1)_L$. It might seem peculiar that a right moving current can generate a left global symmetry but this is precisely the outcome of the calculations in \cite{Hofman:2011zj} and has also been understood from the gravity and string theory perspective in \cite{Compere:2009zj} and \cite{Azeyanagi:2012zd} .

\subsection{The algebra}

If we consider the theory on the plane, the commutators of these operators are given by \cite{Hofman:2011zj}:
\bea
i [T_\xi, T_\zeta] &=& T_{\xi' \zeta-\zeta' \xi} +  \frac{c}{48 \pi} \int dx^- \, (  \xi'' \zeta'-\zeta'' \xi')\label{virplane}\\
i[P_\chi, P_\psi] &=&  \frac{k}{8 \pi} \int dx^- ( \chi'\psi-\psi' \chi )\label{kacplane}\notag\\
i [T_\xi, P_\chi] &=& P_{-\chi' \xi}\label{chargeplane}\notag
\eea
where we have defined 
\be\label{tplane}
T_\xi = -\frac{1}{2\pi}\int dx^- \, \xi(x^-)  T(x^-) \quad P_\chi = -\frac{1}{2\pi}\int dx^- \, \chi(x^-)  P(x^-)
\ee
and $T(x^-)$ and $P(x^-)$ are the usual local operators in the plane. We associate right moving with $x^-$ and left moving with $x^+$.  We furthermore demand that the ground state of the theory is invariant under the action of the global symmetries.

We will be mostly interested in putting this theory on the cylinder. We can describe the cylinder by a complex change of coordinates from the plane. At this point the cylinder theory is Lorentzian with real coordinates and we don't claim that any analytic continuation relates the plane to the cylinder. We will describe the correct relation further on.  Let us then consider the change of coordinates
\be
 x^- = e^{i \phi}. 
\ee
Using the new coordinate
$\phi$ and picking test functions $\xi_n = (x^-)^n=e^{i n\phi}$, we can compute
\bea
\, [L_n, L_m] &=& (n-m) L_{n+m}  + \frac{c}{12} n (n-1)(n+1) \delta_{n+m}\notag\label{JJ}\\
\, [P_n, P_m]  &=&  \frac{k}{2} n \delta_{n+m}\label{LL}\\
 \, [L_n, P_m] &=& -m P_{m+n}\label{LJ}\notag
\eea
with  $L_n = i T_{\xi_{n+1}}$ and $P_n=P_{\chi_n}$. 

In what follows we will perform changes of coordinates to obtain the vacuum energy and charge of a theory and also to manipulate the partition function by modular transformations. With this in mind, we need to track the way the anomalies show up in the transformations of $T$ and $P$. What's more, we need to be able to do this for finite transformations. The procedure is analogous to the one that yields the Schwarzian derivative in standard CFTs.

The commutation relations imply the following infinitesimal transformations of the energy momentum tensor and current:
\begin{eqnarray}
\delta_{\epsilon} T(x^-) &=& - \epsilon(x^-) \partial_- T(x^-) - 2 \partial_- \epsilon(x^-) T(x^-)  -\frac{c}{12} \partial_-^3 \epsilon\\
\delta_\gamma T(x^-) &=& - \partial_- \gamma(x^-) P(x^-)\notag\\
\delta_{\epsilon} P(x^-) &=&-\epsilon(x^-) \partial_- P(x^-) - \partial_- \epsilon(x^-) P(x^-)\notag\\
\delta_\gamma P(x^-) &=&   \frac{k}{2} \partial_{-} \gamma(x^-)\notag
\end{eqnarray}
where we have defined
\be
\delta_{\epsilon +\gamma} = \delta_\epsilon + \delta_\gamma = -i [T_\epsilon, \cdot] - i [P_\gamma, \cdot].
\ee

\subsection{Finite transformations}

Notice that while $T(x^-)$  generates infinitesimal general coordinate transformations in $x^-$, $P(x^-)$ generates gauge transformation in the gauge bundle parameterized by $x^+$. We can think of both these transformations as coordinate transformations of the form
\be
x^- = f(w^-) \quad\quad\quad x^+ = w^+ + g(w^-)
\ee
for arbitrary functions $f,g$.  These reduce to $\delta w^- =  - \epsilon(w^-)$ and $\delta w^+ = - \gamma(\w^-)$ when the transformation is infinitesimal.
By requiring that the finite transformation laws reduce to these infinitesimal versions and that they also compose appropriately we can uniquely fix:
 \be
 P'(w^-) = \frac{\partial x^-}{\partial w^-} \left[ P(x^-) + \frac{k}{2} \frac{\partial w^+}{\partial x^-}\right]\label{jfinite}
\ee
and
 \begin{eqnarray}
 T'(w^-) &=& \left(\frac{\partial x^-}{\partial w^-}\right)^2 \left[ T(x^-) - \frac{c}{12}\left\{\frac{\frac{\partial^3 w^-}{\partial x^{-3}}}{\frac{\partial w^-}{\partial x^-}}- \frac{3}{2} \left( \frac{\frac{\partial^2 w^-}{\partial x^{-2}}}{\frac{\partial w^-}{\partial x^-}}\right)^2 \right\} \right]\nonumber\\
 & & + \frac{\partial x^-}{\partial w^-} \frac{\partial x^+}{\partial w^-} P(x^-) - \frac{k}{4} \left(\frac{\partial x^+}{\partial w^-}\right)^2\label{tfinite} \ .
\end{eqnarray}
Notice that $P(x^-)$ transforms as a $+-$ tensor as one might have imagined. Let us stress the curious appearance of the current anomaly $k$ in the finite transformation law for $T(x^-)$. While this term vanishes to linear order and, thus, does not appear in the algebra (\ref{LL})-(\ref{LJ}), it is unavoidable once $P(x^-)$ mixes with $T(x^-)$.

 Let us now be more specific and specialize this result to a case of interest to us, the mapping from $x^-$ to $\phi$ coordinates. Furthermore, we can add an arbitrary \textit{tilt} $\alpha$.  The change of coordinates is
 \be
x^- = e^{i \phi} \quad\quad\quad x^+ = t +  2 \alpha\,  \phi\label{cylplane} \ .
\ee
We will return to the fact this is complex below, but note that in an ordinary CFT, the analogous mapping from the Lorentzian plane to the Lorentzian cylinder is also a complex coordinate transformation. Using the expressions above we get
\be
P^\alpha(\phi) = i  x^- P(x^-) - k \alpha
\ee
and
\be
T^\alpha(\phi) =- x^{- 2} T(x^-) + \frac{c}{24} + i 2 \alpha x^- P(x^-)  -k \alpha^2
\ee
We can define modes on the cylinder as
\be
P_n^{\alpha} = -\frac{1}{2\pi} \int d\phi \, P^{\alpha}(\phi) \, e^{i n \phi} \quad\quad\quad L_n^{\alpha}  = -\frac{1}{2\pi} \int d\phi \, T^\alpha(\phi) \,e^{i n \phi} 
\ee
In terms of the original modes this is:
\be
P_n^{\alpha} =  P_n + k \alpha \, \delta_n \quad\quad\quad L_n^{\alpha} = L_n + 2 \alpha P_n + (k \alpha^2 -\frac{c}{24}) \delta_n\label{sfcompact}
\ee
We can clearly appreciate that (\ref{sfcompact}) is nothing else than the usual shift proportional to the central charge when doing an exponential mapping combined with a spectral flow transformation given by the tilt parameter $\alpha$. 

A related transformation that will be relevant at finite temperature is
\be
x^- = -\frac{1}{2\pi T_R}e^{-2\pi T_R \phi} \ , \quad x^+ = t + 2 \alpha \phi \ .
\ee
In this case,
\bea\label{finitettrans}
P'(\phi) &=& -2 \pi T_R x^- P(x^-)-k\alpha \\
T'(\phi) &=& (2\pi T_R x^-)^2 T(x^-) - 4 \pi T_R \alpha x^- P(x^-) - k \alpha^2 - \frac{c}{6}\pi^2 T_R^2 \ . \notag
\eea

Finally, further on it will be useful to understand the relation between two sets of coordinates on the cylinder in which we change the size and tilt parameter. This is:
\be
\phi = \frac{\phi'}{\lambda}  \quad\quad t = t' + 2 \frac{\gamma}{\lambda} \, \phi'\label{cc1}  \ .
\ee
Using once again the expression for the finite transformations (\ref{tfinite}) and (\ref{jfinite}) we obtain
\be\label{ccanomcur}
P'(\phi') = \frac{1}{\lambda} \left[P(\phi) - k \gamma\right] \quad\quad T'(\phi') = \left(\frac{1}{\lambda}\right)^2 \left[ T(\phi) + 2 \gamma P(\phi) - k \gamma^2\right] \ .
\ee

In particular, this implies the following relation for the generator of translations with respect to the above coordinates:
\be
Q[\partial_{t'}] = Q[\partial_t]+ k \gamma\ ,  \quad \quad Q[\partial_{\phi'}]= \frac{1}{\lambda}\left(Q[\partial_{\phi}]+ 2 \gamma Q[\partial_t]+ k \gamma^2 \right)  \label{ccanom} \ .
\ee

It is very interesting to interpret the meaning of (\ref{ccanom}). The above formula is just capturing the anomalous terms in the change of coordinates (\ref{cc1}).  Looking at the generators of symmetries we see that the change of coordinates acts like an ordinary transformation of partial derivatives, plus anomalous terms that need to be calculated as above. This will be of importance when discussing the properties of partition functions of these theories under modular transformations.

\subsection{Unitary representations}\label{unit}

Now let us study the unitary representations of the WCFT algebra. If we want to demand that $T^\alpha(\phi)$ and $P^\alpha(\phi)$ should be hermitian we must require
\be
L_{-n} = L_{n}^\dag \quad\quad\quad P_{-n} = P_n^\dag\label{herm} \ .
\ee
We will follow the above convention. Alternatively, one could fix the sign of $k$ and derive the necessary hermiticity conditions compatible with unitarity.

Define primary states as
\be
P_n | p, h \rangle = 0 \quad n>0 \quad\quad L_n | p, h \rangle = 0 \quad n>0
\ee
and
\be
P_0  | p, h \rangle = p | p, h \rangle \quad\quad L_0 | p, h\rangle = h | p, h \rangle \ .
\ee
Positivity of the states $L_{-n}|p,h\rangle$, $P_{-n}|p,h\rangle$, and $P_0|p,h\rangle$ requires
\be
c > 0 \ , \quad k>0 \ , \quad h\geq 0 \ , \quad p \in \mathbb{R}  \ .
\ee
These are not the only constraints coming from unitarity. One can define another energy momentum tensor $T'(x^-)$ where one subtracts the contribution coming from the Sugawara construction of the $U(1)$ current. In terms of modes, define:
\be
L'_n = L_n - \frac{1}{k} \sum_m : P_{n+m} P_{-m}
\ee
where $: :$ indicates normal ordering. It is easily checked that the $L'_n$s are hermitian if the $L_n$s and $P_n$s are. Furthermore they commute with $P_n$s and obey the Virasoro algebra
\be
[L'_n, L'_m] = (n-m) L'_{n+m}  + \frac{c-1}{12} n (n-1)(n+1) \delta_{n+m} \ .\\
\ee
If we now calculate the norm of $L'_{-n} |p, h \rangle$ we find the requirements
\be
h \geq \frac{p^2}{k} \ , \quad c \geq 1 \ .
\ee

In summary, we have: $k>0$, $p \in \mathbb{R}$, $c\geq1$ and $h\geq \frac{p^2}{k}$ in order for representations to be unitary with the conventions (\ref{herm}).

\subsection{States and vacuum energies}
It may happen that on the cylinder, $P_0 \neq 0$ in the vacuum state.  By computing the norms of $L_n$ and $P_n$ descendants, now using the cylinder algebra, we can conclude that in a unitary theory,
\be
L_0 \geq \frac{P_0^2}{k} - \frac{c}{24} \ .
\ee
Therefore, it might seem natural to guess that the charges of the vacuum state saturate this bound and can be parameterized as
\be\label{vaccharge}
P_0^{\alpha, vac} = k \alpha \quad\quad L_0^{\alpha, vac} = k \alpha^2 - \frac{c}{24} \ .
\ee
We will now make a more precise statement, and in the process argue that (\ref{vaccharge}) is true even in non-unitary theories as long as the vacuum state is associated to the unit operator.

So far we have been a bit vague about the connection between the theory on the cylinder and the plane and how to interpret the complex change of coordinates (\ref{cylplane}). 
We want to define states of a WCFT on the cylinder parameterized by the coordinates $\phi$ and $t$. We will do this by analytically continuing $\phi$ at $t=0$. This means that in (\ref{cylplane}), we can replace the coordinate $x^-$ by $z$ in this new complex plane, capping off the Lorentzian cylinder with a Euclidean disk, and insert an operator at the origin. The vacuum charge of the current $P$ can be interpreted as the fact that the holomorphic theory on the $z$ plane has a nontrivial magnetic flux through the origin. This implies directly that we are forced to consider spectral flowed representations, as generated by the tilt in (\ref{cylplane}), fixing the value of $\alpha$ above. The upshot is that states in our WCFT are created by the insertion of spectral flowed operators in a holomorphic (i.e. chiral) CFT containing a current $P$.  The chiral CFT must be very special to ensure locality in the $U(1)$ direction.

An important point is that a general spectral flow transformation does not leave the spectrum invariant. In particular, if we start with a theory with a neutral vacuum state in the plane we expect it to pick up background charges. This is nothing other than the effect of a magnetic flux inside the cylinder. Furthermore, because $P(x^-)$ is the current associated with $x^+$ translations, we expect the spectrum of $P_0$ to be continuous and bounded below. In this case, the spectral flow always maps the vacuum into the vacuum and does not recover the original spectrum for any $\alpha$. This is clearly different from the usual case of a compact $U(1)$.

Having said this we are in a position to calculate the charges of the vacuum state on the cylinder using (\ref{sfcompact}). If we assume that the identity operator $L_0=P_0=0$ in our Euclidean chiral CFT is associated to the vacuum state, we obtain (\ref{vaccharge}) as we predicted.  The spectral flow parameter $\alpha$ is a property of the theory on the cylinder.

\section{Entropy}\label{s:linearensemble}

Consider a WCFT with coordinates $(t,\phi)$ chosen so that the symmetries are 
\be\label{startsym}
\phi \rightarrow f(\phi) , \quad t \rightarrow t - g(\phi) \ .
\ee
Let us put this theory on a circle of unit radius
\be
\phi \sim \phi + 2\pi  \ ,
\ee
at finite temperature and angular potential. It can be shown, using the same symmetries that will prove useful in this section, that an arbitrary choice of  slicing on which we define states can be taken into a circle aligned with the action of $L_0$ in this way.\footnote{There is one degenerate case that constitutes the only exception to this statement. If we try to define states at the $\phi=0$ surface, this amounts to a form of DLCQ, as the $t$ coordinate is always light like. In this case one can show by the same arguments of this section that the entropy is independent of the $P_0$ charge. Curiously, this case might be connected with the understanding of Kerr/CFT \cite{Guica:2008mu}.}

  The partition function at inverse temperature $\beta$ and angular potential $\theta$ is
\be\label{origz}
Z(\beta, \theta) = \Tr\ e^{-\beta P_0 + i \theta L_0}
\ee
where the energy and angular momentum are charges generating the translations
\be
P_0 = Q[\p_t]  \  , \quad L_0 = Q[\p_\phi] \ .
\ee
Thermal correlators are periodic under the complex shift
\be
(t,\phi) \sim (t+i\beta, \phi + \theta) \ .
\ee
We are in Lorentzian signature, so this identification should be interpreted as shorthand for the statement that real-time correlators have the specified periodicity as analytic functions of the coordinates. The arguments in this section can also be made in Euclidean signature, with the same results.

\subsection{Asymptotic Density of states}

Using the Virasoro and Kac-Moody symmetries we can derive a universal formula for the asymptotic density of states of a WCFT, analogous to the Cardy formula in ordinary CFT.  Motivated by the usual derivation of the Cardy formula (reviewed in this language in appendix \ref{app:cardy}), we seek a transformation of the form (\ref{startsym})  that exchanges the thermal cycle with the angular cycle.  This will play the role of a modular transformation. Take the ansatz
\be\label{modtrans}
\phi' = \lambda \phi \ , \quad t' = t -2 \gamma \phi  \ .
\ee
The new periodicities are
\bea
\mbox{thermal:}\quad (t',\phi') &\sim& (t' + i \beta -2 \gamma \theta\ , \ \phi' + \lambda \theta)\\
\mbox{angular:}\quad (t',\phi') &\sim& (t' - 4 \pi \gamma \ , \  \phi' + 2 \pi \lambda)\notag \ .
\eea
Now, choosing
\be
\gamma = \frac{i\beta}{2\theta}\ , \quad \lambda = \frac{2\pi}{\theta} \ ,
\ee
we find the new identifications
\be
(t',\phi') \sim (t', \phi' + 2\pi) \sim (t' + i \beta' \ , \phi' + \theta')
\ee
where
\be\label{newparam}
\theta' = -\frac{4\pi^2}{\theta} \ , \quad \beta' =  \frac{2\pi \beta}{\theta} \ .
\ee
Therefore, the partition function is invariant --- up to an anomaly --- under the `warped modular transformation' (\ref{newparam}). The anomaly arises because (\ref{modtrans}) is not among the global symmetries $SL(2)\times U(1)$. It can be computed by applying (\ref{ccanomcur}), 
\be\label{tchange}
T(\phi) = \frac{4\pi^2}{\theta^2}T'(\phi') - \frac{2\pi i \beta}{\theta^2}P'(\phi') + \frac{k\beta^2}{4\theta^2} \ .
\ee
The operator $ i \int_0^\theta d\phi T(\phi) $, defined by integrating over the thermal cycle of the original torus, becomes the evolution operator on the new torus. Thus the modular transformation, including the anomaly, is
\bea\label{transformz}
Z(\beta,\theta) &=& \Tr\  \exp\left( -\frac{2\pi \beta}{\theta}P_0 - \frac{4\pi^2}{\theta}i L_0 + ik\frac{\beta^2}{4\theta} \right)\\
&=& e^{ik\frac{\beta^2}{4\theta}}Z\left(\frac{2\pi\beta}{\theta}, -\frac{4\pi^2}{\theta}\right) \ . \notag
\eea
We have dropped the primes as the spectrum of the primed operators coincides with the original spectrum.
We can now derive the density of states at small imaginary $\theta$, because in this slowly-rotating regime the warped modular transformation projects the trace onto the state of minimal $L_0$ (for real $\beta$ and $P_0$, the first term is just a phase):
\be
Z(\beta,\theta) \approx \exp\left(-\frac{2\pi \beta}{\theta}P_0^{vac}- \frac{4\pi^2}{\theta}i L_0^{vac} + ik\frac{\beta^2}{4\theta}\right)
\ee
where `vac' means the state with minimal $L_0$ (and we have assumed this state has no macroscopic degeneracy). Obviously this makes sense only if $L_0$ is bounded below, although we will consider another possibility in the following section.\footnote{ Note that if $L_0$ and $P_0$ have real spectra, then for imaginary $\theta$, the original expression for the partition function (\ref{origz}) is manifestly real.  Comparing to (\ref{transformz}), this implies that $P_0$ eigenstates come in positive and negative pairs.  On the other hand if $P_0$ does not come in pairs, then it must have a complex spectrum.}

Using the thermodynamic formula $S = (1-\beta\p_\beta-\theta\p_\theta)\log Z$, the entropy is 
\be\label{entropy}
S = \frac{2\pi i}{\Omega}P_0^{vac} - \frac{8\pi^2}{\beta\Omega}L_0^{vac}\ ,
\ee
where the angular potential is related to the angular velocity $\Omega$ by 
\be
\theta =  i \beta \Omega \ .
\ee
In the microcanonical ensemble,
\be\label{entropymicro}
S = -\frac{4\pi i P_0 P_0^{vac}}{k} + 4\pi\sqrt{-\left(L_0^{vac} - \frac{(P_0^{vac})^2}{k}\right)\left(L_0 - \frac{P_0^2}{k}\right)} \ .
\ee
To go any further, we would need to determine $L_0^{vac}, P_0^{vac}$, which may depend on the particular theory.  We have argued in (\ref{vaccharge}) that under reasonable assumptions, the vacuum state of a WCFT can be usefully parameterized by spectral flowing from the trivial vacuum.  Plugging in 
\be\label{vaccharge2}
P_0^{vac} = q \ , \quad L_0^{vac} =\frac{q^2}{k}-\frac{c}{24} \ ,
\ee
the entropy becomes
\be\label{smicrocardy}
S =-4\pi i \frac{q P_0}{k} + 2 \pi \sqrt{\frac{c}{6}\left(L_0 - \frac{P_0^2}{k}\right)} \ .
\ee
This Cardy-like formula is one of our main results.  Below, we will compare to WCFTs defined holographically, and use this formula to reproduce the black hole entropy. 
 The entropy formula is valid in the slowly-rotating regime
\be
c \gg \beta \Omega \ , \quad\quad \frac{\Delta_{gap}}{\beta\Omega} \gg 1 \  ,
\ee
where $\Delta_{gap}$ is the dimension of $L_0$ where the theory starts to have a large number of operators.  One might expect $\Delta_{gap} \sim 1/c$ in a typical theory, which gives the sufficient condition $\frac{1}{\beta\Omega}\gg c \gtrsim 1$, but more generally, the precise domain of validity depends on the spectrum of $L_0$. 

In the derivation of (\ref{smicrocardy}), we assumed that $L_0$ is bounded below, but we did not assume hermiticity of $P_0$ or that the theory is unitary.  If $P_0$ is hermitian, then $q$ must vanish so the first term does not appear.  If the theory is not unitary, then strictly speaking this is not an entropy since the partition function has negative contributions, but $S$ still measures the asymptotic behavior of $Z$.

\subsection{$SL(2,Z)$}

The warped modular transformation together with other symmetries of the warped CFT actually generate $SL(2,Z)$.  To see this, define
\be
\tau = \frac{\theta}{2\pi}
\ee
and note that we have the transformations
\bea
S: \quad \tau \rightarrow -\frac{1}{\tau}\\
T: \quad \tau \rightarrow \tau + 1 \notag
\eea
where $S$ is the warped modular transformation and $T$ comes from adding the angular circle to the thermal circle. Together these generate $SL(2,Z)$. Under $S$, the partition function transforms as
\be
 Z\left(-\frac{1}{\tau}, \frac{\beta}{\tau}\right) = e^{-ik\frac{\beta^2}{8\pi\tau}}Z(\tau, \beta)  \ .
\ee
This is the transformation rule for a weak Jacobi form, familiar in the context of superconformal field theory from the transformation of the elliptic genus, see e.g. \cite{Kraus:2006nb}. 

\section{Other ensembles and nonlocal algebras}\label{s:quadraticensemble}

In this section, we consider a modified algebra where the central term in the $U(1)$ algebra is charge-dependent,
\bea\label{dalgebra}
\, [\tL_n  \ , \tL_m] &=& (n-m)\tL_{n+m} + \frac{c}{12}(n^3-n)\delta_{n+m}\notag\\
\, [\tL_n \ , \tP_m ] &=& -m\tP_{m+n} + m \tP_0\delta_{n+m}\label{ll2}\\
\, [ \tP_n \ , \tP_m] &=& 2 n \tP_0 \delta_{m+n} \notag \ .
\eea
The motivation will become apparent when we reach the holographic examples in sections \ref{s:tmg} and \ref{s:strings}.  This form of the algebra also makes it easier to connect our Cardy-like formula (\ref{smicrocardy}) to the actual Cardy formula in an ordinary CFT. It is related to the original algebra (\ref{LL}) by redefining charges as
\be\label{jmap}
\tP_n = \frac{2}{k}P_0 P_n - \frac{1}{k}P_0^2 \delta_n \ , \quad \quad \tL_n = L_n - \frac{2}{k}P_0 P_n + \frac{1}{k}P_0^2\delta_n \ .
\ee
For states with vanishing $P_{n\neq 0}$, this amounts to a nonlocal reparameterization of the theory where the time coordinate is rescaled by the total energy, 
\be
x^+ = \frac{kt}{2P_0} + \phi  \ , \quad\quad x^- = \phi \ .
\ee
Indeed, (\ref{dalgebra}) cannot be written as the variations of local currents. Notice, that the above algebra looks like spectral flow by an amount proportional to $P_0$ as far at the $\tL_n$'s go, but it involves a rescaling of the current as was pointed out above. This construction is very reminiscent of the Sugawara construction of a Virasoro algebra from quadratic combinations of Kac-Moody generators. Indeed, if one looks at (\ref{ll2}) for the cases where the anomalous terms contribute (i.e. $n+m=0$), the above algebra coincides with two copies of commuting Virasoros. On the other hand for $n+m \neq 0$ it agrees with the Virasoro-Kac-Moody algebra.
The reason we chose this algebra is because it appears naturally from gravity where the Killing vectors of the metric yield the classical $U(1)$ contribution to the commutators, while the quantum anomalies of the associated charges make it look like a Virasoro. Notice that if we had picked the Sugawara representation, classical looking terms in the commutators appear as a consequence of the anomalous $U(1)$ current contribution and can't be associated with the algebra of Killing vectors. 

Let us now analyze this case along the lines of sections \ref{s:primer} and \ref{s:linearensemble}.  Consider the infinitesimal tilt,
\be
\delta x^+ =  -\frac{\delta \gamma}{2} x^- \ .
\ee
The zero modes transform as
\be
\delta \tL_0 = 0\ , \quad\quad \delta \tP_0 =  \tP_0 \delta \gamma  \ 
\ee

Note that the anomaly in $[\tL_n,\tP_m]$ has completely canceled the classical term, leaving $\tL_0$ invariant under $U(1)$ transformations. 
If we also include the rescaling $\phi' = \lambda \phi$, the generators of translations transform as:
\be\label{tildecc}
\tQ[\partial_{+'}]= e^{\gamma}\tQ[\partial_{+}] \ , \quad\quad \tQ[\partial_{-'}]= \frac{\tQ[\partial_-] }{\lambda} \ .
\ee
This is analogous to equation (\ref{ccanom}), modified to include a charge-dependent level.  It differs in an important way: under a shift in $x^+$, the charge $\tP_0$ is rescaled.  Unlike (\ref{ccanom}), this does not take the form of a simple tensor transformation plus anomalous shifts.  This means that we must be careful in how we interpret coordinate transformations in this theory, as the anomaly plays a crucial role. In fact, the (active) finite transformation of $\tP_0$ mimics the (passive) coordinate transformation $t \rightarrow e^{\gamma} t$.  We will see that this second scaling leads to CFT-like behavior of the partition function.

Now consider the theory at finite temperature in the ensemble
\be\label{qens}
Z(\beta_L, \beta_R) = \Tr \ e^{-\beta_L \tP_0 -\beta_R \tL_0} \ ,
\ee
on the circle
\be
(x^+\ , x^-) \ \sim \ (x^+ + 2\pi \ , \ x^- + 2\pi) \ .
\ee
Notice that the identification on the circle is implemented by the operator $e^{2\pi i ( \tP_0 + \tL_0)}$ which, in terms of the original charges (\ref{LL}-\ref{LJ}) is nothing else than $e^{2\pi i L_0}$. We are considering the same circle. The ensemble is different, however, as in terms of the original charges
\be
Z = \Tr\ \exp\left[-\beta_L \left(\frac{P_0^2}{k}\right) - \beta_R \left(L_0 - \frac{P_0^2}{k}\right)\right]\ .
\ee

We would like to repeat the steps of section \ref{s:linearensemble} leading to the entropy formula, taking care of the anomaly.  The answer should be the same, because the microcanonical entropy does not depend on a choice of ensemble, but this derivation will be valid where the previous one was not, including when $L_0$ is not bounded below.  The strategy, phrased in operator language, is to find a symmetry transformation that turns the initial evolution operator in (\ref{qens}) into an angular generator of length $2\pi$.  Then, the old angular generator is used as the new evolution operator.  The first step is achieved by choosing
\be
\lambda = -\frac{2\pi i}{\beta_R} \ , \quad\quad e^{-\gamma} = -\frac{2\pi i}{\beta_L} \ .
\ee
Now the generator that enforces  the angular identification, integrated over the (original) thermal circle becomes the evolution operator
\be
e^{-\frac{4 \pi^2}{\beta_L} \tP_0 -\frac{4 \pi^2}{\beta_R} \tL_0}\,.
\ee
Therefore, the partition function can be rewritten on the transformed torus as
\be
Z(\beta_L,\beta_R) = Z(\frac{4 \pi^2}{\beta_L},\frac{4 \pi^2}{\beta_R}) \,.\label{mod}
\ee
This result is exactly as one would have obtained in the usual CFT case! Notice however that all currents are right moving in this theory and the classical algebra contains a Kac-Moody part instead of a left moving Virasoro.

To finalize the argument and as before, we can take the small $\beta_{R,L}$ limit and project the right hand side of (\ref{mod}) onto the vacuum state,
\be
Z(\beta_L,\beta_R) \approx \exp\left(-\frac{4\pi^2}{\beta_L} \tP_0^{vac}  -\frac{4\pi^2}{\beta_R} \tL_0^{vac}\right) \ .
\ee
From this expression the entropy can be calculated and the Cardy result can be obtained
\be
S = -8 \pi^2 \left(\frac{\tP_0^{vac}}{\beta_L}+\frac{\tL_0^{vac}}{\beta_R}\right) \,.
\ee
In terms of charges this is:
\be
S= 4\pi \sqrt{-\tP_0^{vac} \tP_0} + 4\pi \sqrt{-\tL_0^{vac} \tL_0} \,. \label{cardy2}
\ee

This agrees with (\ref{entropymicro}). This is to be expected, as the degeneracy of states in the Hilbert space is independent of the particular ensemble we are considering. Notice, however, that to project onto the vacuum we only need to have $\tL_0$ and $\tP_0$ bounded below. Therefore, this derivation applies to cases where the $L_0$ operator considered in the previous section is unbounded. We will see this is the case in the gravitational theory. 

Finally, let us conclude this section with the following remark. It is inevitable to notice the similarity of (\ref{cardy2}) and the usual Cardy formula. In order to make this concrete we define the following quantities:
\be
c_R = -24 \tL_0^{vac}, \quad\quad c_L = - 24 \tP_0^{vac},
\ee
so we recover the familiar form
\be
S= 2\pi \sqrt{\frac{c_L}{6} \tP_0} + 2\pi \sqrt{\frac{c_R}{6} \tL_0} .\label{cardy3}
\ee
In terms of the vacuum values of the charges $L_0$ and $P_0$ displayed in (\ref{vaccharge}) we obtain
\be
c_R = c, \quad\quad c_L = - 24 \frac{q^2}{k}.
\ee
Notice that while $c_R$ is connected with the actual central charge of the algebra, $c_L$ is just the amount of spectral flow.\footnote{It is true, however, that once a current algebra is found, one can build a twisted energy momentum tensor through the Sugawara construction. In this case the twisting can shift the vacuum value of the zero mode while changing the central charge of the algebra. It is then a matter of choice whether $c_L$ appears or not in the algebra of generators. The entropy formula (\ref{cardy3}) is, of course, invariant under this twisting.} In the next section we'll find that $c_L$ and $c_R$ are precisely the parameters one naturally finds in gravitational theories with warped black hole solutions.
\section{A holographic example: Topologically Massive Gravity}\label{s:tmg}

The action of topologically massive gravity (TMG) in three dimensions is \cite{Deser:1982vy, Deser:1981wh}
\be\label{tmgaction}
S_{TMG} = \frac{1}{16 \pi}\int d^3 x \sqrt{-g}(R + 2) - \frac{1}{96\pi \nu}\int d^3 x\sqrt{-g}\epsilon^{\lambda\mu\nu}\Gamma^{r}_{\lambda\sigma}\left(\p_\mu\Gamma^{\sigma}_{r\nu} + \frac{2}{3}\Gamma^\sigma_{\mu\tau}\Gamma^r_{\nu r}\right)\ .
\ee
Any solution of Einstein gravity is also a solution of TMG, but the gravitational Chern-Simons term allows for interesting new classes of solutions.  These include warped AdS (WAdS) and associated `warped black holes.'  Warping is a deformation that changes the asymptotics and reduces the isometry group to $SL(2,R) \times U(1)$, so these backgrounds do not fall under the usual AdS/CFT correspondence.  The boundary conditions can be chosen so that the symmetries enhance to Virasoro plus a $U(1)$ Kac-Moody algebra generating diffeomorphisms near the boundary in the sense of Brown and Henneaux, suggesting that the holographic dual is a warped CFT. 

Warped CFTs with a microscopic field theory definition are not known except in some limiting cases, so potential holographic examples are a good testing ground for the technology developed above.  In this section we will show that warped CFT reproduces the thermodynamics of the warped black holes.  Under the assumption that the dual theory exists and has a spectrum that satisfies a property analogous to the gap condition in AdS/CFT, the density of states in the QFT accounts for the Bekenstein-Hawking entropy of the warped black holes.

\subsection{Warped AdS}
 We take $\nu > 0$, and generally follow the notation and terminology of \cite{Anninos:2008fx}.\footnote{The sign convention for the CS action in (\ref{tmgaction}) has been flipped compared to \cite{Anninos:2008fx}. Our choice leads to $L_0 = + Q[\p_\phi]$ below, allowing for a simpler comparison to the WCFT.} Several solutions of TMG with $SL(2,R) \times U(1)$ local isometries are of interest to us:

\subsubsection*{Global spacelike WAdS}

The metric is
\be\label{globalsl}
ds^2 = \frac{1}{\nu^2 + 3}\left[-\cosh^2\sigma d\tau^2 + d\sigma^2 + \frac{4\nu^2}{\nu^2+3}(du + \sinh\sigma d\tau)^2\right]\ ,
\ee
with the coordinates unrestricted.  When $\nu=1$, the isometries enhance to $SL(2,R) \times SL(2,R)$, and this becomes AdS$_3$ written as a Hopf fibration over AdS$_2$.  Generally, the fiber is warped; for $\nu <1$ it is squashed, and for $\nu >1$ it is stretched.  All four isometries are globally preserved and the spacetime is geodesically complete \cite{Anninos:2009zi}.

\subsubsection*{Timelike WAdS}

This can be written in similar global coordinates by taking $u \rightarrow i \tau$, $\tau \rightarrow i u$, or as
\be\label{timelikemetric}
ds^2 = -dt^2 + \frac{dr^2}{r\big((\nu^2+3)r + 4\big)} - 2 \nu r dt d\phi + \frac{r}{4}\big(3(1-\nu^2)r + 4)d\phi^2 \ ,
\ee
with $\phi \sim \phi + 2\pi$. These coordinates cover the global spacetime.  For $\nu > 1$, there are closed timelike curves at large $r$.

\subsubsection*{Poincar\'{e} spacelike WAdS}
The metric is
\be\label{poincaremetric}
ds^2 = dt^2 + \frac{dr^2}{r^2(\nu^2+3)} - 2\nu r dt d\phi + \frac{3}{4}(\nu^2-1)r^2 d\phi^2
\ee
with $\phi$ unidentified.  This covers a patch of the global spacetime (\ref{globalsl}).

\subsubsection*{Spacelike stretched black holes}
Finally, for $\nu >1$ there are the warped black holes.  These are locally spacelike stretched WAdS (\ref{globalsl}), but differ globally by an identification that breaks the isometries to $U(1) \times U(1)$.  Thus these are the warped analogues of the BTZ black holes in AdS$_3$.  The metric in Schwarzschild coordinates is
\begin{align}\label{blackhole}
ds^2 &= dt^2 + \frac{dr^2}{(\nu^2+3)(r-r_+)(r-r_-)} - \left( 2\nu r - \sqrt{r_+ r_-(\nu^2+3)}\right)dtd\phi\\
&\quad \quad +\frac{r}{4}\left(3(\nu^2-1)r + (\nu^2+3)(r_++r_-) - 4\nu\sqrt{r_+r_-(\nu^2+3)}\right)d\phi^2\notag
\end{align}
where $\phi \sim \phi + 2\pi$. When $r_+ = r_- = 0$, the charges vanish and this becomes an identification of Poincar\'{e} WAdS.

\subsection{Asymptotic symmetries and thermodynamics}
It is possible to impose boundary conditions in WAdS that allow for a right-moving Virasoro algebra \textit{or} a left-moving Virasoro algebra in the asymptotic symmetries, but no consistent boundary conditions have been found that allow two Virasoro algebras simultaneously.  We will impose the boundary conditions of \cite{Compere:2008cv}, which extend $SL(2,R) \times U(1)$ to a Virasoro-Kac-Moody $U(1)$ algebra. Let us emphasize that this is a choice that defines the theory under consideration, and there may be other consistent choices with different interpretations.

The generators of asymptotic diffeomorphisms allowed by the boundary conditions are \cite{Compere:2008cv}
\bea \label{AKV}
\zeta_n &=& e^{i n \phi}\p_\phi  - i n r e^{i n \phi}\p_r \\
\chi_n &=& e^{i n \phi}\p_t  \ .\notag
\eea
These satisfy the Lie bracket algebra
\be
i [\zeta_n\ , \ \zeta_m ]_{\mbox{\small Lie}} = (n-m)\zeta_{n+m} \ , \quad\quad i [\zeta_n\ ,\  \chi_m]_{\mbox{\small Lie}} = -m\chi_{n+m} \ .
\ee
The corresponding charges (see appendix \ref{app:charges}), 
\be
L_n = Q[\zeta_n] \ , \quad \quad P_n = Q[\chi_n] \ ,
\ee
satisfy the Virasoro-Kac-Moody $U(1)$ algebra (\ref{LL}) under Dirac brackets, with central extensions
\be
c = \frac{5\nu^2 + 3}{\nu(\nu^2+3)} , \quad \quad k = -\frac{\nu^2+3}{6\nu} \ .
\ee
By rescaling $t$ one can also rescale the level $k$, but charges are rescaled accordingly so that expressions of the form $P P /k$ are unchanged.

The charges and thermodynamics of the black hole take a simple form expressed in terms of $c,k$, and the parameters
\begin{align}
T_L &= \frac{\nu^2+3}{8\pi}\left(r_+ + r_- - \frac{1}{\nu}\sqrt{r_+r_-(\nu^2+3)}\right)\\
T_R &= \frac{\nu^2+3}{8\pi}\left(r_+ - r_-\right) \ .
\end{align}
For now, these are just useful parameterizations of the black hole and have no obvious interpretation as temperatures.  The black hole mass $\mathcal{M}$ and angular momentum $\mathcal{L}$, including contributions from the Chern-Simons term, are
\begin{align}\label{bhcharges}
\mathcal{M} & := Q[\p_t] =  \frac{\pi}{3} T_L\\
\mathcal{L} &:= -Q[\p_\phi] = -\frac{1}{k}\mathcal{M}^2 - \frac{\pi^2}{6}c\, T_R^2\notag \ .
\end{align}
The inverse Hawking temperature and angular potential are
\begin{align}
\beta &= -\frac{2\pi}{3k}(1 + T_L/T_R) \\
 \beta\Omega &= 1/T_R\notag \ .
\end{align}
The black hole entropy (which also includes a Chern-Simons contribution) is
\be
S_{bh} = \frac{\pi}{3\Omega} + \frac{\pi^2}{3\beta \Omega}(c + \frac{2}{3k})\ .
\ee
In the microcanonical ensemble, 
\be\label{sbhmicro}
S_{bh}  = -\frac{2\pi}{3k} \mathcal{M} + 2 \pi \sqrt{\frac{c}{6}\left(-\mathcal{L} - \frac{\mathcal{M}^2}{k}\right)} \ .
\ee
In the rest of this section, the goal is to reproduce this formula from warped CFT.

\subsection{The ensemble}\label{ensTMG}
To compare the black hole thermodynamics to a warped CFT, we must decide what ensemble to use.  In other words, what is the black hole dual to? The answer should be a thermal state, but different types of thermal states were considered in sections \ref{s:linearensemble} and \ref{s:quadraticensemble}.  Charge-dependent coordinate changes suggest different ensembles, since nonlinear redefinitions of the charges lead to inequivalent partition functions.

In the bulk, the thermal properties of the black hole are summarized by the complex coordinate identifications
\be\label{bhid}
(t,\phi) \sim (t + i \beta, \phi + i \beta \Omega) \ .
\ee
The zero modes of the algebra are
\be
P_0 = \mathcal{M} \ , \quad \quad L_0 = -\mathcal{L} \ ,
\ee
so this suggests the thermal ensemble studied in section (\ref{s:linearensemble}),  $\tr \ e^{-\beta P_0 + i \theta L_0}$.  However this is problematic since $L_0$ is not bounded below, and we will argue for a different interpretation.

To derive the ensemble, we can use the fact that different black holes are related to each other by a coordinate transformation. To clarify the logic, we first review the analogous argument for BTZ black holes in the AdS$_3$/CFT$_2$ correspondence, made in \cite{Maldacena:1998bw}: The coordinate transformation from Poincar\'{e} AdS$_3$ to the BTZ black hole is $w^{\pm} \sim e^{2\pi T_{\pm}(\phi\pm t)}$ near the boundary. The thermal identification on $t,\phi$ is trivial in the $w^\pm$ plane, so the black hole corresponds to the Minkowski vacuum in $w^\pm$ coordinates.  The exponential coordinate transformation covers the Rindler wedge, producing a thermal state in $t,\phi$ coordinates; therefore, the black hole is dual to a thermal ensemble $\tr\,e^{-\beta \mathcal{M} - \beta \Omega \mathcal{L}}$.

Now we return to the warped black holes and repeat the same steps.  Starting with coordinates $(t,r,\phi)$ on the warped black hole (\ref{blackhole}), let 
\bea
r' &=& \sqrt{(r-r_+)(r-r_-)}e^{2\pi T_R \phi}\\
\phi' &=& \frac{2}{3+\nu^2}\frac{r_++r_--2r}{(r_+-r_-)\sqrt{(r-r_+)(r-r_-)}} e^{-2\pi T_R \phi}\notag\\
t' &=& t + \frac{2}{k}\mathcal{M} \phi + \frac{\nu}{3+\nu^2}\log\left(r-r_+\over r-r_-\right)\ . \notag
\eea
The metric in the primed coordinates is Poincar\'{e} WAdS (\ref{poincaremetric}), i.e., the $L_0 = P_0 = 0$ black hole. Near the boundary, this coordinate transformation that creates a black hole is
\bea\label{exptobh}
\phi' &=&-\frac{1}{2\pi T_R} e^{-2\pi T_R \phi} + O(1/r^2)\\
t' &=& t +\frac{2}{k}\mathcal{M} \phi + O(1/r) \ .\notag
\eea
The energy and angular momentum measured in $\phi,t$ coordinates come from anomalous transformations of $L_0$ and $P_0$.  Applying (\ref{finitettrans}) with $\alpha = \mathcal{M}/k$, the  anomalies produce
\begin{align}
L_0 &= \frac{c}{6}\pi^2 T_R^2 + k\alpha^2 = -\mathcal{L}\\
P_0 &= k\alpha = \mathcal{M}
\end{align}
in agreement with (\ref{bhcharges}).

In the $t',\phi'$ plane, the thermal identification (\ref{bhid}) acts as 
\be
(t', \phi') \sim (t' + i \beta_0, \phi')
\ee
where
\be
\beta_0 = -\frac{2\pi}{3k} \ .
\ee
From this we conclude that the black hole ensemble is defined by starting in the plane at temperature $\beta_0$, and performing the coordinate change (\ref{exptobh}).  To understand the resulting state, define
\be
\phi'' = -\frac{1}{2\pi T_R}e^{-2\pi T_R \phi } \ , \quad t'' = \frac{1}{2\pi T_L}\exp\left(2\pi T_L \left(\frac{k}{2\mathcal{M}}t +\phi\right) \right)\ .
\ee
The black hole corresponds to the Minkowski vacuum in the $\phi'', t''$ plane. The exponential coordinate changes are just the usual map to Rindler space, so this produces a thermal state, but the appearance of $\mathcal{M}$ in the transformation means that $t$ is an inconvenient coordinate to define the ensemble. In terms of the more natural coordinates
\be
t_R = \phi , \quad \quad t_L = \frac{k}{2\mathcal{M}}t + \phi \  ,
\ee
the exponential map turns on temperatures $T_{L,R}$ conjugate to the charges $Q[\p_{L,R}]$. The infinitesimal charges obey
\be
\delta Q[\p_L] = \frac{2\mathcal{M}}{k}\delta \mathcal{M} , \quad\quad \delta Q[\p_R] = -\delta\mathcal{L} -\frac{2\mathcal{M}}{k}\delta\mathcal{M} \ .
\ee
Integrating,
\be
Q[\p_L] = \frac{P_0^2}{k} \ , \quad \quad Q[\p_R] = L_0 - \frac{P_0^2}{k} \ .
\ee
Therefore, the black hole is dual to the thermal ensemble 
\be\label{bhensemble}
Z_{bh} = \Tr \ \exp\left[ -\beta_R \left(L_0 -\frac{P_0^2}{k}\right) - \beta_L \frac{P_0^2}{k}\right] \ .
\ee
This is the quadratic ensemble studied in field theory terms in section \ref{s:quadraticensemble}, with
\be
\beta_{L,R} = T_{L,R}^{-1} \ .
\ee

\subsection{BTZ-like coordinates}\label{btzlikecoords}
The coordinates $t_{L,R}$ that appeared naturally in the derivation of the ensemble are actually coordinates on a deformed BTZ black hole. Define new coordinates $(t_b, \phi_b, r_b)$ by
\begin{align}
\phi_b &-\frac{t_b}{\ell_b} = t_R = \phi\\
 \phi_b &+ \frac{t_b}{\ell_b}= t_L = \frac{k}{2\mathcal{M}}t +\phi \notag \\
r_b^2 &= 3 \mathcal{M}\left(2r - \frac{1}{\nu}\sqrt{r_+ r_-(\nu^2+3)}\right)  +4 \ell_b J_{\rm BTZ}\notag
\end{align}
where $J_{\rm BTZ}$ is defined below and 
\be
\ell_b^2 = \frac{4}{3+\nu^2} \ .
\ee
The resulting metric can be written in the form
\be\label{btzform}
ds^2 = ds^2_{BTZ}  + \frac{1}{48}(\nu^2-1) \xi_\mu \xi_\nu dx_b^\mu dx_b^\nu \ .
\ee
The first term here is the BTZ black hole in AdS$_3$ of radius $\ell_b$,
\be
ds^2_{BTZ} = \left(8M_{BTZ} - \frac{r_b^2}{\ell_b^2}\right)dt_b^2 + \frac{dr_b^2}{-8M_{BTZ} + \frac{r_b^2}{\ell_b^2} + \frac{16J_{BTZ}^2}{r_b^2}} -8 J_{BTZ} dt_b d\phi_b + r_b^2 d\phi_b^2 \ ,
\ee
where the BTZ mass and angular momentum are related to the warped black hole parameters by
\bea\label{btzchargerelation}
\mathcal{M} &=& \frac{1}{6}\sqrt{8(M_{BTZ} - J_{BTZ}/\ell_b)} \\
\mathcal{L} &=& -\frac{M_{BTZ}}{3\nu} -\frac{1+ 3\nu^2}{\nu(\nu^2+3)}\frac{J_{BTZ}}{\ell_b}\ .\notag
\eea
The second term in (\ref{btzform}) is a deformation by the Killing vector
\be
\xi = \frac{1}{\mathcal{M}}\left(\ell_b \p_{t_b} + \p_{\phi_b}\right) \ ,
\ee
with the index on $\xi^\mu$ lowered using the undeformed BTZ metric. These coordinates have the advantage that the ensemble is defined with potentials conjugate to $\p_{t_b}$ and $\p_{\phi_b}$. However, they have the disadvantage that the leading asymptotics of the metric are $\mathcal{M}$-dependent, for instance for $J_{BTZ} = 0$, 
\be
\frac{4}{3 (\nu^2-1)} ( ds^2 - ds^2_{BTZ} ) =  \frac{(r_b^2 - 8\ell_b^2 M_{BTZ} )^2}{8\ell_b^2 M_{BTZ}} dt_b^2 + 2 \ell_b( r_b^2 - \frac{r_b^4}{8\ell_b^2 M_{BTZ}}) dt_b d\phi_b + \frac{r_b^4 }{8M_{BTZ}} d\phi_b^2.
\ee
This complicates the task of defining charges and computing asymptotic symmetries as compared to Schwarzschild coordinates, but the result is simply given by the map (\ref{jmap}). 

\subsection{Entropy from warped CFT}

Finally, we are ready to compare the entropy of warped CFT to the black hole entropy (\ref{sbhmicro}).   The warped CFT entropy formula (\ref{entropymicro}) requires the charges of the `ground state', so we must identify the appropriate state in TMG.  The choice of the quadratic ensemble does not affect the microcanonical formula for the entropy, but it means that the ground state is defined by minimizing the shifted charge $L_0 - \frac{P_0^2}{k}$.

As described in section \ref{s:primer}, we expect the ground state to have global isometries $SL(2,R) \times U(1)$ and charges
\be\label{vacrel}
P_0^{vac} = q \ , \quad \quad L_0^{vac} = -\frac{c}{24} + \frac{q^2}{k} \ ,
\ee
where the vacuum charge $q$ (or rather the invariant combination $q^2/k$) is a parameter of the theory. Therefore we seek a smooth solution of TMG with these properties.  Given the relation to the BTZ black hole described above, a natural guess for the ground state is to take the deformation of global AdS rather than BTZ.  The global AdS metric is $ds^2_{BTZ}$ with $M_{BTZ} = -\frac{1}{8}$, $J_{BTZ} = 0$.  Plugging this into (\ref{btzchargerelation}) indeed gives the relations (\ref{vacrel}) with
\be\label{bulkgroundstate}
q_{TMG} = \mathcal{M}^{vac} =  -\frac{i}{6} \ , \quad L_0^{vac}= -\mathcal{L}^{vac} = -\frac{c}{24} - \frac{1}{36 k} \ .
\ee
Furthermore, the full metric (\ref{btzform}) is smooth for this value of the parameters --- in fact, it is timelike warped AdS (\ref{timelikemetric}) --- and minimizes $L_0 - P_0^2/k$ among known smooth solutions.  Therefore this is the ground state.  Plugging into the entropy formula (\ref{entropymicro}) and comparing to (\ref{sbhmicro}), we find
\be
S_{bh} = S_{wcft} \ .
\ee

Some comments are in order. In the original $(t,r,\phi)$ coordinates, the warped black hole metric is complex in the ground state (\ref{bulkgroundstate}), which corresponds to 
\be
r_+ = -\frac{4i}{3+\nu^2} \ , \quad r_- = 0 \ .
\ee
Nonetheless it has a natural interpretation: continuing $r \to i r$, $t \to -i t$ gives the global timelike warped AdS metric (\ref{timelikemetric}).  The complex metric in $(t,r,\phi)$ coordinates reflects the $\mathcal{M}$-dependent rescalings necessary to define the ensemble and change to BTZ-like coordinates.  In BTZ-like coordinates the metric remains real everywhere in phase space.  The fact that the vacuum has complex $P_0$ is related to the appearance of closed timelike curves (CTCs) in the bulk. This matches nicely with the discussion in section \ref{unit} and indicates the theory is not unitary. This is, however, exactly what is needed to match the entropy.

The warped CFT entropy was derived in the limit $\theta \rightarrow 0$, but correctly matches the classical entropy for arbitrary black holes.  This implies that the spectrum of a warped CFT dual to TMG must be  special, to ensure that the entropy formula applies outside its generic regime of validity.  This is analogous to the fact that in AdS$_3$/CFT$_2$, the Cardy formula matches the black hole entropy even for black holes well outside the generic Cardy regime (see \cite{Keller:2011xi} for a discussion). In string theory realizations, this is achieved by having a large gap in operator dimensions in the CFT. A similar condition is sufficient in the warped case.

\section{An example in string theory}\label{s:strings}

We now turn to an embedding of warped black holes in string theory. Using a series of transformations that relate this solution to an ordinary BTZ black hole, it was argued in \cite{ElShowk:2011cm, Song:2011sr}  (see also \cite{Bena:2012wc,Azeyanagi:2012zd,Anninos:2008qb}) that the dual field theory is the IR limit of the dipole-deformed D1-D5 field theory.  In principle, this defines the dual theory, but in practice, little is known about the IR limit after the dipole deformation.  We will take a complementary approach, using the symmetries to motivate the conjecture that the warped black holes in string theory have a warped CFT description.  As evidence for this conjecture, we show that the density of states in warped CFT reproduces the entropy of the warped black holes.  There are some peculiar features to this construction that we will not attempt to address --- the black hole has closed timelike curves unless the angle is unwrapped, and for related reasons the interpretation of what states are being counted is unclear --- so this should not be considered a full microscopic derivation of the entropy, but it is nonetheless suggestive.

\subsection{Lightlike dipole background}
The bulk theory is a consistent truncation of IIB supergravity to six dimensions with four scalars and two 2-forms \cite{Duff:1998cr}.  With a constant dilaton, and three scalars that play no role in our discussion set to zero, the relevant part of the action is
\be
S = \frac{1}{16\pi}\int d^6x\sqrt{-g}\left(R - \frac{1}{12}e^{-2\phi}H^2 - \frac{1}{12}F^2 \right) \ ,
\ee
with 3-form field strengths $H=dB$, $F=dA$. We will focus on the finite-temperature lightlike dipole background (following the notation in \cite{ElShowk:2011cm}),
\bea\label{strsol}
e^{\phi}ds^2 &=& \frac{T_+^2dy^2}{1+\lambda^2T_+^2} + \frac{2rdy dt}{1+\lambda^2 T_+^2} + dt^2\left(T_-^2 - \frac{\lambda^2 r^2}{1+\lambda^2T_+^2}\right)\\
& & \quad  + \frac{dr^2}{4(r^2-T_+^2T_-^2)} + \frac{1}{4(1+\lambda^2 T_+^2)}(d\psi + \cos\theta d\phi)^2 + \frac{1}{4}(d\theta^2 + \sin^2\theta d\phi^2)\notag\\
B &=& \frac{\lambda }{2(1+\lambda^2 T_+^2)}(T_+^2 dy + r dt)\wedge (d\psi + \cos\theta d\phi) \notag\\
A &=& r dy \wedge dt + \frac{1}{4}\cos\theta d\psi \wedge d\phi \notag\\
e^{-2\phi} &=& 1 + \lambda^2 T_+^2 \notag \ .
\eea
This can be obtained from BTZ$\times S^3$ by performing a TsT transformation with shift $\lambda$ .  The metric is squashed AdS$_3$ times squashed $S^3$. Defining
\be
t = x + \tau, \quad y = x - \tau  \ ,
\ee
the angular identification of the original BTZ is along $x$.  We previously limited ourselves to the stretched case $\nu>1$ because squashed black holes have CTCs.  Therefore in this case we will unwrap the circle and compute charges per unit length in the $x$-direction. 

The inverse Hawking temperature $\beta$ and angular velocity $\Omega$ are related to the parameters $T_{\pm}$ by
\be
T_\pm = \frac{\pi}{\beta(1\pm\Omega)} \ ,
\ee
and the entropy per unit length is
\be\label{strentropy}
S = \frac{\pi^3 }{\beta(1-\Omega^2)} \ .
\ee
The asymptotic symmetries include Virasoro + $\hat{U}(1)$ generated by
\bea
\zeta_n &=& e^{i n t}\left(\p_t - i n r \p_r \right)\\
\chi_n &=& e^{int}\p_y \notag \ ,
\eea
with corresponding charges $\tL_n = Q[\zeta_n]$, $\tP_n = Q[\chi_n]$, and zero modes
\be
\tL_0 = \frac{\pi}{4}T_-^2 \ , \quad \tP_0 = -\frac{\pi}{4}T_+^2 \ .
\ee
The level and central charge, computed by the standard method, are
\be\label{strlevel1}
c = \frac{3\pi}{2} \ , \quad \tilde{k} = -\pi T_+^2 \ .
\ee
Note that we have set $\ell=G=1$.

\subsection{Entropy}

The asymptotic algebra suggests that there is a warped CFT description.  We will now show that the warped CFT can also be used to reproduce the entropy.  In (\ref{strlevel1}), $\tP_0$ appears on the right-hand side of the algebra, so the coordinates (\ref{strsol}) are similar to the coordinates of section \ref{btzlikecoords}, where the leading terms in the metric include the charges.  To eliminate this complication, define
\be\label{stringuv}
u = t \ , \quad v = T_+ (y-t) \ .
\ee
In these coordinates, the algebra takes the standard form (\ref{LL}), with
\be
k = -\pi \ , \quad P_0 = -\frac{\pi}{2}T_+ \ , \quad L_0 = \tL_0 + \tP_0\ .
\ee
To derive the entropy from WCFT, we temporarily identify $x \sim x + 2\pi$ (later we can put the theory back on the plane; the same step would be necessary to derive the entropy of an ordinary CFT per unit length).  The angular and thermal identifications are then
\be\label{uvid}
(u,v) \sim (u + 2\pi, v) \sim (u + \theta+i\beta, v  - 2 i T_+ \beta)
\ee
where $\theta = -i\beta\Omega$.  The shift in (\ref{stringuv}) was chosen so that the first identification acts only on the $SL(2,R)$ coordinate, since this tilt was assumed in the derivation of the entropy formula in field theory.

Applying the WCFT entropy formula (\ref{entropy}) using the potentials in (\ref{uvid}), we find
\be
S = -\frac{8\pi^2}{\beta(1-\Omega)}L_{0}^{vac}  + \frac{4\pi^2}{\beta(1-\Omega^2)}iP_{0}^{vac}\label{entstr} \ .
\ee
The correct bulk entropy (\ref{strentropy}) is obtained for
\be
L_{0}^{vac} = 0 \ , \quad P_{0}^{vac} = -\frac{i}{4}\pi \ .
\ee
The solution with these charges is the timelike vacuum, just as for TMG discussed in section \ref{s:tmg}. Therefore, assuming this geometry contributes to the partition function, the WCFT entropy formula agrees with the Bekenstein-Hawking entropy.

It is important to stress that (\ref{strentropy}) is obtained by calculating an entropy per unit length in a timelike direction (although it is spacelike in the original BTZ black hole before performing the TsT transformation).  This is the same quantity that was reproduced from the ordinary Cardy formula in \cite{Song:2011sr}.  How such a quantity should be interpreted in terms of states in the quantum theory is unclear, so the imaginary charge and negative level that we encountered do not contradict unitarity.

\section{Discussion}

In this work we have studied two dimensional field theories that, while lacking Lorentz invariance, posses enough structure so that their global symmetries can be extended to an infinite dimensional local algebra \cite{Hofman:2011zj}. We have shown that this algebra constrains the asymptotic density of states of these so called WCFTs in a similar fashion to the standard Cardy argument \cite{Cardy:1986ie} for CFTs. The former theories have a form of modular invariance which can be used to obtain concise expressions for their entropy which resemble the well known Cardy formula. This is the main result of our work.

Given the lack of examples in a field theory context, we decided to turn to some known proposals for holographic duals. In this context, we used our result to explain the entropy of warped black holes in TMG. It is shown that the Bekenstein-Hawking entropy exactly matches our field theory prediction. It is worth mentioning that we are able to do this without invoking the presence of a hidden second Virasoro algebra. This gives some evidence that the dual field theory to this gravitational setup is a WCFT, while the more familiar CFT structure needs not be present.

In passing we have shown that it does not seem to be possible to have a fully consistent (i.e. unitary) quantum theory of gravity for the warped solutions of TMG with $\nu > 1$ and the standard choice of boundary conditions. While it is true that space-like stretched black hole solutions do not present CTCs, it seems one is forced to include the time-like deformed solutions in the spectrum if the symmetries of the theory are to be preserved. In this cases CTCs appear and unitarity is lost.  This may explain why, in microscopic constructions, only squashed AdS has appeared.

We also studied the better behaved example of lightlike dipole deformed backgrounds in string theory. These solutions can be obtained by TsT transformations of the usual AdS$_3 \times S^3$ solution in Type IIB string theory and, therefore, have a consistent UV completion. The theory admits black string type solutions where the angular direction of the BTZ parent background is unwrapped. Nevertheless, a formula for the Bekenstein-Hawking entropy per unit length of these solutions is known and we show it agrees with the predictions of WCFT. Once again, no reference to a second hidden Virasoro algebra is needed to prove the result.

It is interesting to point out, nonetheless, that while WCFTs posses a Virasoro-Kac-Moody algebra, there is a sense in which the current mimics the presence of a second Virasoro algebra. It was shown in section \ref{ensTMG} that the ensemble that naturally describes the gravitational setups  leads us to a consider a nonlocal algebra, described in section \ref{s:quadraticensemble}, that shares some properties of a second scaling symmetry. Even more so, we can mention the following curious fact. Let us focus on the nonlocal U(1) algebra given by
\be\label{jjf}
[ \tP_n \ , \tP_m] = 2 n \tP_0 \delta_{m+n} .
\ee
This algebra can be obtained by a nonlocal contraction of a Virasoro algebra. Let us define
\be
\tP_0 = \epsilon^2 L_0,\quad\quad \tP_n = \epsilon L_n \quad \textrm{for} \quad n \neq 0 .
\ee
Then, the commutators (\ref{jjf}) can be obtained from the Virasoro algebra by taking $\epsilon \rightarrow 0$.

There is, of course, another related way in which a second Virasoro algebra appears. This is simply by considering the Sugawara construction of Virasoro generators from a Kac-Moody algebra. At the level of zero modes, this is identical to our non-local algebra, but the full Sugawara generators seem difficult to realize as asymptotic symmetries.  Furthermore, for a $U(1)$ algebra this leads to a Virasoro central charge $c = 1$ (although this can be remedied by twisting, at the expense of adding a new free parameter \cite{Blagojevic:2009ek}).  

Whether any of these remarks is connected with the presence of a second hidden Virasoro algebra remains an open question, which is of particular interest for the understanding of the Kerr/CFT correspondence. Although we have not focused on this particular case, it is of clear interest to extend the ideas discussed in this work to the study of Kerr and other black holes. An interesting piece of information in this direction was mentioned in section \ref{s:linearensemble}. The near horizon extremal Kerr (or NHEK) geometry  at fixed polar angle is warped AdS$_3$, but with the angular identification purely in the $U(1)$ direction.  This is a degenerate case where the spatial circle cannot be rotated to align with the $SL(2,R)$ zero mode, so the results of section \ref{s:linearensemble} do not apply directly.  This case in TMG, known as the self-dual solution, is included in our entropy formula, but only as a limit where one temperature is taken to zero.  

This line of thought could also be generalized to maximally rotating black holes in four-dimensional de Sitter space \cite{Anninos:2009yc}, whose near-horizon geometry contains a warped dS$_3$ factor instead of WAdS$_3$ but nevertheless exhibits a similar symmetry structure \cite{Anninos:2009jt, Anninos:2011vd}.

More generally, every extremal black hole has $SL(2,R)\times U(1)$ isometries in the near horizon.  This comes from the  appearance of an AdS$_2$ factor; it would be interesting to compare our construction to gravity in AdS$_2$, along the lines of \cite{Hartman:2008dq,Castro:2009jf,Castro:2010vi}. Even away from extremality, a large class of black holes in flat space have entropies that resemble the Cardy formula.  This fact has been partially understood for Kerr from a hidden conformal symmetry in the linearized equations of motion \cite{Castro:2010fd}, but without some notion of a modular transformation, the entropy remains a puzzle.  Every black hole has a torus, of course, so the proliferation of Cardy-like entropy formulae hints that perhaps symmetries can be used to swap the thermal and angular cycles in other cases as well.

Another useful test of holographic dualities, especially because it can be applied without a microscopic definition of the field theory, is the comparison of scattering amplitudes \cite{Das:1996wn} or quasinormal modes \cite{Birmingham:2001pj}. The scattering cross sections of various fields on Kerr black holes have been matched to a CFT \cite{Bredberg:2009pv} (see also the review \cite{Compere:2012jk}), but in the CFT approach this required adding a current algebra and imposing a constraint of the form $L_0 = J_0$. Similar analyses have been performed for the WAdS$_3$ \cite{Oh:2008tc,Kao:2009fh, Chen:2009hg, Chen:2009cg, Yao:2011kf}  and WdS$_3$ black holes \cite{Chen:2011dc}. The symmetries of WCFT also constrain correlation functions \cite{Hofman:2011zj}, so it would be interesting to revisit these computations.

One interesting aspect of WCFTs is that they force us to consider $T(x^-)$ and $P(x^-)$ on equal footing. As mentioned in the introduction, maybe this is a toy example that might help us understand higher spin theories. If one is to understand the modular properties of partition functions of these more sophisticated theories this feature can not be overlooked.

Lastly let us comment on the importance of understanding the meaning of holography in other asymptotic spacetimes. While the most interesting cases are dS and flat space in higher dimensions, which are sure to be different, historically much has been learnt from the study of the very symmetric cases associated with two dimensional field theories. Hopefully, other structures as powerful as the one described here can be uncovered for other cases of interest.

\section*{Acknowledgements}

It is a pleasure to thank the Stanford Institute for Theoretical Physics and the Kavli Institute for Theoretical Physics where conversations got this work started, as well as the conference `Cosmology and Complexity'. We have also benefitted greatly from discussions with Dionysios Anninos, Alejandra Castro, Monica Guica, Sean Hartnoll, Juan Maldacena, Gim-Seng Ng, Wei Song, and Philippe Spindel, and especially thank Andrew Strominger for discussions and for comments on a draft. D.M.H. would like to thank the Center for the Fundamental Laws of Nature at Harvard University for support. T.H. acknowledges support by the U.S. Department of Energy, grant DE-FG02-90ER40542, and the Corning Glass Works Foundation Fellowship Fund. S.D. acknowledges support from the Fundamental Laws Initiative, of the Center for the Fundamental Laws of Nature, Harvard University, and from Wallonie-Bruxelles International.

\appendix

\section{Derivation of the Cardy Formula in Ordinary CFT}\label{app:cardy}
The Cardy formula is a universal result on the high-temperature density of states in an ordinary 2d CFT \cite{Cardy:1986ie}.  In this appendix, we review the derivation in the same Lorentzian language used in section \ref{s:linearensemble}.   (See also \cite{Carlip:1998qw} for a clear discussion on the role of vacuum charges in the Cardy formula.) The partition function is defined as
\be
Z(\beta,\theta) = \Tr\ e^{-\beta H - i \theta J}  = \Tr\ q^{L_0}\bar{q}^{\bar{L}_0} \ , 
\ee
where
\be
q = e^{2\pi i \tau}  \ ,\quad 2\pi \tau = \theta + i \beta \ .
\ee
and $2\pi\bar{\tau} = \theta - i \beta$.
Finite temperature correlators are periodic under
\be
(t,\phi) \sim (t, \phi + 2 \pi) \sim (t + i\beta, \phi + \theta) \ .
\ee
Defining 
\be
x^\pm = \phi \pm t \ ,
\ee
the periodicities are
\be
(x^+, x^-) \sim (x^+ + 2 \pi, x^- +  2 \pi) \sim (x^+ + 2\pi \tau, x^- + 2\pi \bar{\tau})
\ee
In ordinary CFT, the symmetries allow independent rescalings of $x^\pm$.  Therefore we seek a transformation of the form
\be
x^+ \rightarrow \lambda_+ x^+ \ , \quad x^- \rightarrow \lambda_- x^-
\ee
that interchanges the thermal and spatial cycles.  This can be achieved by setting 
\be
\lambda_+ = -1/\tau \ , \quad \lambda_- = -1/\bar{\tau} \ ,
\ee
so that the new periodicities are
\be
(x^+,x^-) \sim (x^+ + 2 \pi \tau', x^- + 2 \pi \bar{\tau}') \sim (x^+ - 2 \pi, x^- - 2 \pi)
\ee
with
\be
\tau' = -1/\tau \ .
\ee
Therefore we have derived invariance of the partition function under the $S$ modular transformation
\be
Z(\tau, \bar\tau) = Z(-1/\tau, -1/\bar\tau) \ ,
\ee
or
\be
Z(\beta,\theta) = Z\left(4\pi^2\frac{\beta}{\theta^2+\beta^2}, \ -4\pi^2\frac{\theta}{\theta^2 + \beta^2}\right)
\ee
At high temperatures, this projects the trace onto the state of minimal $H$, ie the vacuum, so
\be
Z(\beta,\theta) \approx \exp(-\frac{4\pi^2\beta}{\theta^2+\beta^2} H_{vac}  + \frac{4\pi^2\theta}{\theta^2+\beta^2}iJ_{vac}) \ .
\ee
This implies the microcanonical entropy
\be
S_{CFT} = 2\pi\sqrt{-(H_{vac} + J_{vac})(H + J)} + 2\pi\sqrt{-(H_{vac}-J_{vac})(H-J)} \ .
\ee
This is as far as we can go in a completely general CFT.  (This version of the Cardy formula, written in terms of vacuum charges rather than central charges, has proved useful in holographic applications where the groundstate is not ordinary AdS$_3$   \cite{Correa:2010hf,Correa:2011dt}.) In a unitary CFT, the unit operator on the plane provides the vacuum state on the cylinder, so we can compute the vacuum charges from the conformal transformation to the cylinder,
\be
H_{vac} = L_0 + \bar{L}_0 = -\frac{c_L + c_R}{24} \ , \quad J_{vac} = L_0 - \bar{L}_0 = -\frac{c_L - c_R}{24} \ .
\ee
Plugging in gives the usual Cardy formula,
\be
S_{CFT} = 2\pi\sqrt{\frac{c_L}{6}L_0} + 2\pi\sqrt{\frac{c_R}{6}\bar{L}_0} \ .
\ee

\section{Charges in Topologically Massive Gravity}\label{app:charges}
In this appendix we collect formulae from \cite{Abbott:1981ff,Iyer:1994ys,Bouchareb:2007yx,Compere:2008cv} for the conserved charges in topologically massive gravity.  In the covariant formalism, the infinitesimal charge associated to an asymptotic Killing vector $\zeta$ is
\be\label{chargevar}
\delta Q[\zeta] = \frac{1}{16\pi}\int k[\zeta; h, g] \ ,
\ee
integrated over the boundary of a fixed-time surface. Here $h_{\mu\nu} = \delta g_{\mu\nu}$ is a linearized solution to the equations of motion, and the integrand can be written
\be
k[\zeta; h, g] = \epsilon_{\mu\nu\rho}\left(k_{grav}^{\mu\nu}[\zeta; h,g] + k_{cs}^{\mu\nu}[\zeta; h,g] \right)dx^\rho \ .
\ee
The Einstein contribution is
\begin{align}
k_{grav}^{\mu\nu}[\zeta; h, g] = & \zeta^\nu(D^\mu h - D_\sigma h^{\mu\sigma}) + \zeta_\sigma D^\nu h^{\mu\sigma} + \frac{1}{2} h D^\nu \zeta^\mu \\
& - h^{\nu\sigma}D_\sigma\zeta^\mu + \frac{1}{2}h^{\sigma \nu}(D^\mu\zeta_\sigma + D_\sigma \zeta^\mu)\notag
\end{align}
and the Chern-Simons term contributes
\begin{align}
k_{cs}^{\mu\nu}[\zeta; h, g] &= \frac{1}{3\nu}k_{grav}^{\mu\nu}[\eta; h, g] - \frac{1}{6\nu}\zeta_\lambda\left(2 \epsilon^{\mu\nu\rho}\delta(G^\lambda_{\ \rho}) - \epsilon^{\mu\nu\lambda}\delta G\right)\\
& \quad\quad +\frac{1}{6\nu}\epsilon^{\mu\nu\rho}\left(\zeta_\rho h^{\lambda \sigma}G_{\sigma\lambda} + \frac{1}{2}h(\zeta_\sigma G^\sigma_{\ \rho} + \frac{1}{2}\zeta_\rho R)\right) \ , \notag
\end{align}
where $\eta^\mu = \half\epsilon^{\mu\nu\rho}D_\nu\zeta_\rho$. (We have discarded a `supplemental term' in the CS contribution \cite{Compere:2008cv} that vanishes for Killing vectors and does not contribute to any of the charges computed in this paper.) Finite charges are computed by integrating the variation (\ref{chargevar}) from one solution to another.

\end{document}